# Magnetic microscopy and simulation of strain-mediated control of magnetization in Ni/PMN-PT nanostructures

Ian Gilbert[1], Andres C. Chavez[2], Daniel T. Pierce[1], John Unguris[1], Wei-Yang Sun[2], Cheng-Yen Liang[2], and Gregory P. Carman[2]

1. Center for Nanoscale Science and Technology, National Institute of Standards and Technology, Gaithersburg, MD 20899
2. Department of Mechanical and Aerospace Engineering, University of California, Los Angeles, CA 90095

Strain-mediated thin film multiferroics comprising piezoelectric/ferromagnetic heterostructures enable the electrical manipulation of magnetization with much greater efficiency than other methods; however, the investigation of nanostructures fabricated from these materials is limited. Here we characterize ferromagnetic Ni nanostructures grown on a ferroelectric PMN-PT substrate using scanning electron microscopy with polarization analysis (SEMPA) and micromagnetic simulations. The magnetization of the Ni nanostructures can be controlled with a combination of sample geometry and applied electric field, which strains the ferroelectric substrate and changes the magnetization via magnetoelastic coupling. We evaluate two types of simulations of ferromagnetic nanostructures on strained ferroelectric substrates: conventional micromagnetic simulations including a simple uniaxial strain, and coupled micromagnetic-elastodynamic simulations. Both simulations qualitatively capture the response of the magnetization changes produced by the applied strain, with the coupled solution providing more accurate representation.

One of the primary goals of the field of spintronics is to electrically control magnetization in a reliable and efficient manner[1-4]. Many methods for manipulating magnetization electrically have been explored experimentally. The coupling between ferroelectric and ferromagnetic order parameters in single-phase multiferroic systems such as $BiFeO_3$[5,6] is one avenue by which electrical control of magnetism may be achieved[7-9]. Another is to utilize the spin transfer torque exerted on an ultrathin magnetic free layer by a spin-polarized current[10,11] from a ferromagnetic polarizer or a pure spin current generated by spin-orbit torque in a heavy metal layer[12,13]. Finally, one can take advantage of the magnetoelastic coupling in a piezoelectric/ferromagnet heterostructure, an approach that is currently gaining significant attention[14-17]. An applied electric field generates a strain in the piezoelectric layer via the converse piezoelectric effect, and this strain changes the ferromagnetic layer's easy axis through magnetostriction. Simulations suggest that strain in a piezoelectric can switch the moment of a nanomagnet while dissipating less than 1.5 aJ, making this approach the most promising from the perspective of energy efficiency[18,19].

Here we use scanning electron microscopy with polarization analysis (SEMPA)[20] to investigate the magnetization changes generated in Ni films and submicron disks by strain in the underlying ferroelectric $[Pb(Mg_{1/3}Nb_{2/3})O_3]_{0.68}[PbTiO_3]_{0.32}$ (PMN-PT) substrate produced by an applied electric field. SEMPA images before and during the application of the electric field allow us to precisely determine the induced changes to the three-dimensional vector magnetization with resolution in the tens of nanometers. The magnetoelectric coupling has a strong effect on vortex magnetization patterns in the Ni disks: it can compress a vortex into two antiparallel domains that point along the easy axis defined by the strain, or it can completely remove the vortex core, converting the disk to a single-domain state. We then use these images to quantitatively evaluate the accuracy of two types of micromagnetic simulations: one which treats the strain with a simple uniform uniaxial anisotropy, and another that fully couples elastodynamic equations with the Landau-Lifshitz-Gilbert (LLG) equation to capture the effects of the local structure of the strain on the magnetization. While both models produce satisfactory results, the fully-coupled elastodynamic and micromagnetic simulation produces the most accurate results.

The sample geometry considered here is illustrated in Figure 1a. Submicron Ni disks were patterned on a 10 mm × 10 mm × 0.5 mm single crystal PMN-PT (011) substrate. The [100] in-plane crystallographic axis of the PMN-PT substrate is aligned with the sample's y-direction while the [01$\bar{1}$] in-plane crystallographic axis is aligned with the x-direction. Planar electrodes are deposited on the PMN-PT's top (30 nm Pt) and bottom (30 nm Ta) surfaces by electron beam evaporation. The disks (diameters 300 nm to 1000 nm) are patterned with electron beam lithography on the top (Pt) surface using a double layer of PMMA/MMA e-beam resist followed by electron beam evaporation of Ti(5 nm)/Ni(12 nm) and liftoff. The PMN-PT substrate was poled with a 0.8 MV/m electric field before the Ti/Ni film was deposited. Applying a post-poling electric field of 0.8 MV/m produces anisotropic in-plane strain[21]. The differences in strain between $E = 0$ MV/m and $E = 0.8$ MV/m are ε$_{yy}$ = 1200 µm/m and ε$_{xx}$ = -3200 µm/m, as shown in Figure 1b for a similar substrate. The large strain jump at 0.6 MV/m is due to an electric field-induced phase transformation from the rhombohedral phase to the orthorhombic phase that is strongly dependent upon the PMN-PT composition.

Nanoscale imaging of the magnetization of the patterned structures was performed using scanning electron microscopy with polarization analysis (SEMPA). The initial magnetization configuration was set by applying a 120 mT magnetic field in the +x direction. The surface was cleaned in situ with Ar$^+$ ion beam etching monitored with Auger electron spectroscopy. Following cleaning, a few monolayers of Fe were evaporated onto the sample, a standard technique used to increase the spin polarization measured by SEMPA without altering the structure of the underlying magnetization[22]. We first studied the effect of an applied electric field on the magnetization of large Ni rectangles. Figure 2 shows SEMPA images of a corner of one of these rectangles. On the Ni portion of the sample, the magnetization of the Fe layer follows that of the underlying Ni, whereas on the surrounding substrate, the Fe magnetization forms its own domain structure. Without an applied electric field (Figure 2b), the Ni is mostly magnetized in the ±y direction (determined by the rectangle's shape anisotropy), while the Fe on the surrounding substrate is mostly magnetized in the ±x direction. The differences in orientation may be due to residual strain in the substrate that sets the initial Fe magnetization configuration during growth. When an electric field of 0.8 MV/m is applied (Figure 2c), the substrate is strained, and the Ni magnetization rotates to point in the ±x direction while the Fe rotates to point in the ±y direction. In Figures. 2e and 2f, polar plots show the distribution of magnetization directions extracted from Figures 2b and 2c, respectively. The opposite preferred magnetization axes of the Fe and Ni regions of the sample are due to the opposite signs of the magnetostriction coefficients for these two materials, i.e. Fe has a positive magnetostriction coefficient while Ni has a negative magnetostriction coefficient. The electric-field-induced anisotropic strain from the PMN-PT substrate produces an easy axis in the Fe along the ±y direction and an easy axis in the Ni along the ±x direction.

The nickel disks were also measured with SEMPA (e.g., Figures 3a-3d). Consistent with the micromagnetic simulations described below, the large disks exhibit a vortex magnetization pattern, while the small disks exhibit uniform magnetization. The diameter at which the crossover from vortex to single domain occurs (approximately 500 nm) is not completely consistent due to the metastability of each state and the edge roughness in the individual disks[23]. Using SEMPA, however, we can directly image the magnetization of each individual disk in its initial configuration and in the presence of an applied electric field, so the exact magnetization changes due to the magnetoelastic coupling with the substrate can be resolved on the nanoscale.

We focus on Ni disks initially magnetized in a vortex configuration, which allows us to probe the effect of strain on magnetization in all in-plane directions. In Figure 3a-3d, we present SEMPA images of 400 and 600 nm Ni disks before and during the application of a 0.8 MV/m electric field. The 400 nm disk initially contains an off-center vortex (a), but the strain-induced anisotropy removes the vortex and rotates the disk's magnetization to point in the +x direction (b). The 600 nm disk initially contains a vortex located in the disk's center (c). Upon the application of the electric field, the regions of the magnetization parallel to the x-axis grow, while the regions of magnetization parallel to the y-axis shrink (d), as one would expect given the negative magnetostriction of Ni. The vortex core is not removed, but the strain in the piezoelectric substrate effectively compresses the vortex into two antiparallel domains. Note that because the strain is uniaxial rather than unidirectional, the disk does not enter a uniformly magnetized configuration.

We use these results to validate two methods of modeling magnetization changes induced by strain. First, two nickel disks, 400 nm and 600 nm in diameter, were simulated by coupling the LLG equation for micromagnetics with the mechanical strains and stresses via the equations of elastodynamics[24] (assuming small elastic deformations and linear elasticity). This iterative modeling approach mathematically couples the magnetization and displacement states to fully capture the interdependent nonuniform distribution of strain and magnetization in the Ni disks. The nickel disk was discretized using tetrahedral elements with a size on the order of the exchange length of nickel (8.5 nm). The ground state of each disk was determined by starting the system from a randomly oriented magnetization state and allowing it to settle into a stable configuration. A compressive strain of 3200 µm/m and a tensile strain of 1200 µm/m were applied along the x and y-directions of the substrate, respectively[21]. The material properties used for the nickel disk were $M_s = 4.8 \times 10^5$ A/m, $A_{ex} = 1.05 \times 10^{-11}$ J/m, α=0.08, $\lambda_{100} = -46 \times 10^{-6}$, $\lambda_{111} = -24 \times 10^{-6}$, $c_{11} = 2.5 \times 10^{11}$ N/m², $c_{12} = 1.6 \times 10^{11}$ N/m², $c_{44} = 1.18 \times 10^{11}$ N/m². The evaporated Ni film grain size is on the order of 3 nm, so magnetocrystalline anisotropy is negligible due to its relation to exchange length. Details of the numerical solution used have been previously presnted[25]. This model has also been experimentally validated for ring structures on the thick PMN-PT substrates[15] as well as on nanoscale structures[26].

The results of the fully-coupled simulations are presented in Figure 3e-h. Unlike the 400 nm disk in Figure 3a, the 400 nm disk modeled here has a largely uniformly-magnetized ground state, though the magnetization does rotate at the disk's edge. The overall magnetization rotates by 90° upon the application of strain. Again, we note that the edge roughness of the experimental disks and the metastability of the vortex state prevent the completely consistent experimental generation of a uniformly-magnetized ground state. The magnetization of the 600 nm Ni disk, initially in a circular vortex configuration, is compressed somewhat by the application of strain, qualitatively consistent with the SEMPA images.

To further validate these simulations, we also modeled the Ni disks using MuMax3, a standard micromagnetic simulation package[27]. We modeled a 400 nm and a 600 nm diameter Ni disk first without and then with a 150 mT uniaxial anisotropy to capture the effects of strain, where the effective field was calculated using $H = 3\lambda_s c_{44}(\varepsilon_{xx} - \varepsilon_{yy})/M_s$. These results are presented in Figure 3i-l. When the uniaxial anisotropy is imposed, the magnetization rotates to the +x direction, consistent with Figure 3b. The 600 nm disk shows a vortex ground state in the micromagnetic simulation and, consistent with the SEMPA data shown in Figure 3d, the vortex in the 600 nm disk is compressed by the strain.

In order to more quantitatively compare the SEMPA images and the results of the two simulations, we extracted several line cuts of the magnetization angle $\phi$ (defined in the inset of Fig. 4a) taken in circles about the center of the 600 nm disk. Representative line cuts with a 225 nm radius are displayed in Figure 4 as a function of angular position $\theta$ on the disk. The most important item to note from these plots is that all three cuts show the same functional form for both the unstrained and strained configuration. In the unstrained state, the disk's magnetization is a circularly-symmetric vortex, and the magnetization angle varies linearly with angular position on the disk (a). In the strained case, as a consequence of the negative magnetoelastic response of the disk to the applied strain, magnetization in the x-direction (compressive direction) is favored at the expense of magnetization in the y-direction (tensile direction), causing the two step-like features seen in (b). The plot shows that the magnetization in both the top and bottom halves of the disk is more uniform than in the unstrained state, i.e., the slope of the $\phi$ vs $\theta$ curve is shallower on the steps in (b) relative to (a). The differences between the two simulations and the fits to the SEMPA data are shown in Figures 4c and d for the unstrained and strained cases, respectively. While (as expected) both simulations accurately represent the Ni magnetization in the unstrained state, the coupled elastodynamic-LLG simulation more correctly represents the magnetization in the strained state. Note in particular that the slope of the $\phi$ vs. $\theta$ curve in Figure 4b is more accurately captured by the elastodynamic-LLG simulation. The differences between the two simulations might be more pronounced if a stronger magnetoelastic material such as Terfenol-D (Tb$_{0.3}$Dy$_{0.7}$Fe$_{1.92}$) were used in the study. The constant offset between that simulation and the SEMPA data in Figure 4b is likely due to pinning at the irregular edge of the disk[28], which can be seen in Figure 3c as well.

We use SEMPA to directly image the vector magnetization of Ni structures before and after straining the underlying ferroelectric substrate and show that the magnetoelectric coupling allows the magnetization patterns to be manipulated. In particular, we demonstrate that the strain produces a uniaxial easy axis for the magnetization. For a disk initially magnetized in a vortex, the magnetization configuration is either eliminated from the disk or compressed into two antiparallel domains. These results can be successfully modeled both with basic micromagnetic simulations incorporating a spatially-uniform uniaxial anisotropy to model the strain as well as by fully-coupled micromagnetic elastodynamic simulations, though the latter more accurately captures the local effects of strain. For stronger magnetoelastic materials or operation near instabilities it may be necessary to use a fully coupled solution rather than simply adding a spatially-uniform magnetic anisotropy. We anticipate that the techniques described here will be useful in the design of devices utilizing strain to control magnetization[14-17].

**Acknowledgements:** This project was supported by the National Institute of Standards and Technology, Center for Nanoscale Science and Technology under project number R13.0004.04. I.G. acknowledges support from the National Research Council's Research Associateship Program. Work at UCLA was supported in part by FAME, one of six centers of STARnet, a Semiconductor Research Corporation program sponsored by MARCO and DARPA.

**Figures**

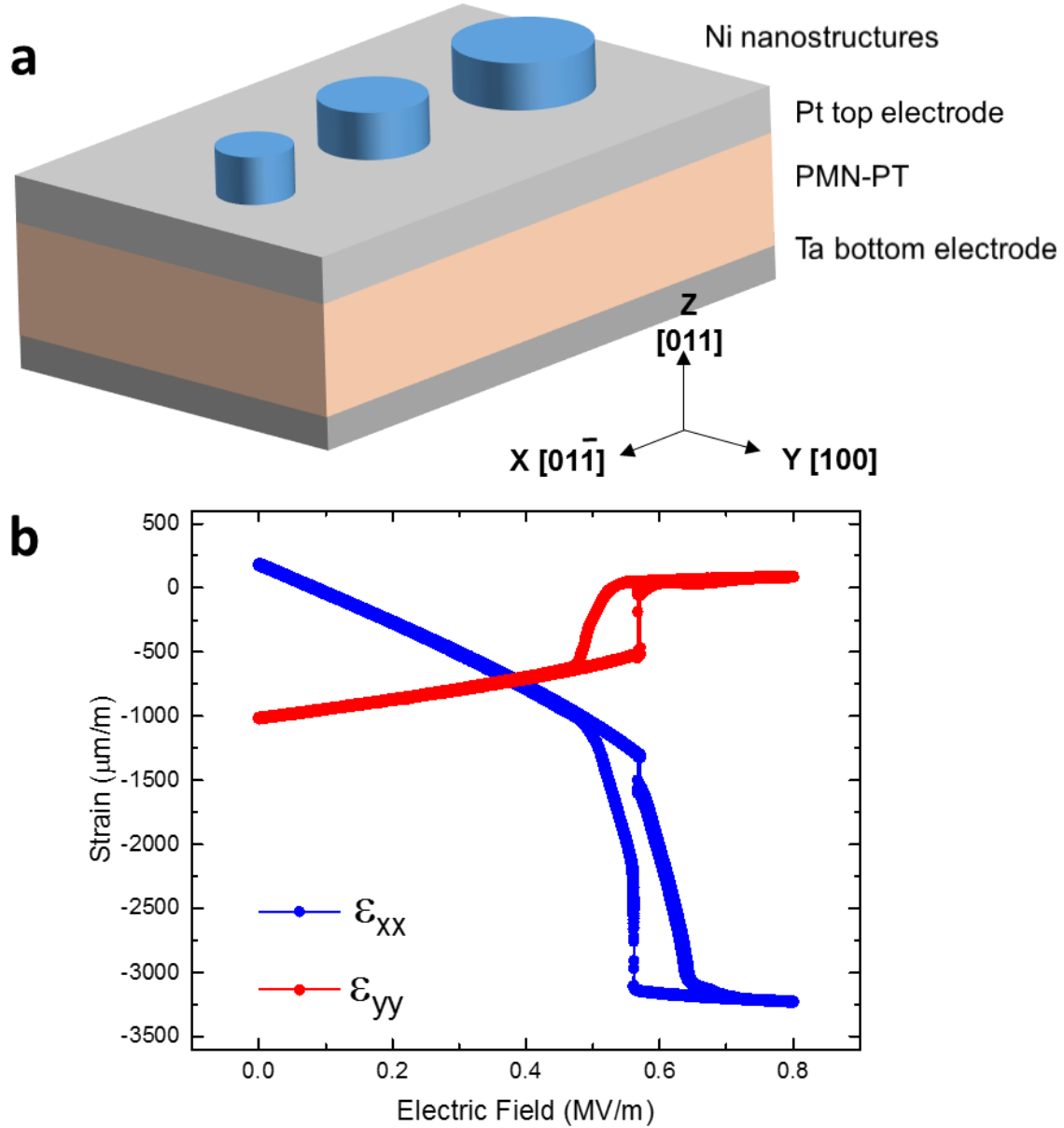

**Figure 1.** (a) Diagram of the sample geometry used in this work. Pt and Ta electrodes on either side of a 500 μm thick PMN-PT substrate are used to apply an electric field in the z direction, which leads to a compressive stress in the x direction and a tensile stress in the y direction. This stress changes the configuration of the magnetization of the Ni disks fabricated on the substrate. (b) Plot of strain in a similar PMN-PT substrate as a function of applied electric field. A large change in strain in both directions occurs around 0.6 MV/m. Uncertainties are derived from uncertainty of the instrument and are less than 5%.

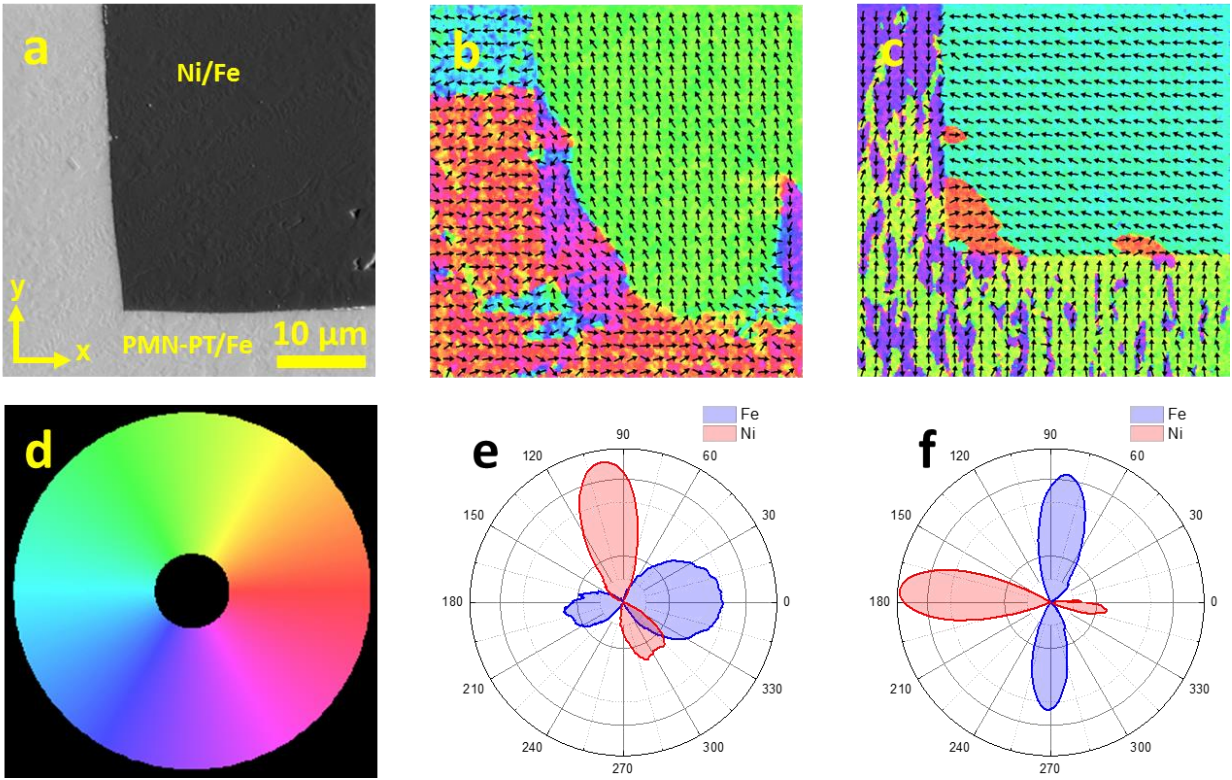

**Figure 2.** (a) A scanning electron micrograph showing the corner of a large Ni rectangle on a PMN-PT substrate. The entire area (substrate and Ni) is covered with a few monolayers of Fe to improve magnetic contrast in SEMPA. (b) SEMPA image showing the magnetization of the Ni rectangle as well as the thin layer of Fe on the PMN-PT substrate without an applied electric field. Panel (c) shows the same area after a strain in the substrate is generated by a 0.8 MV/m electric field. The magnetization directions in (b) and (c) are given by the color wheel in (d). This color scale is also used for all subsequent magnetization images in this work. Panels (e) and (f): Polar plots showing the distribution of magnetization directions present in images (b) and (c), respectively. The blue portions represent the Fe magnetization, and the red portions represent the Ni magnetization. The strain produced by the applied electric field rotates the magnetization 90° and also slightly reduces the spread of the distribution of magnetization angles.

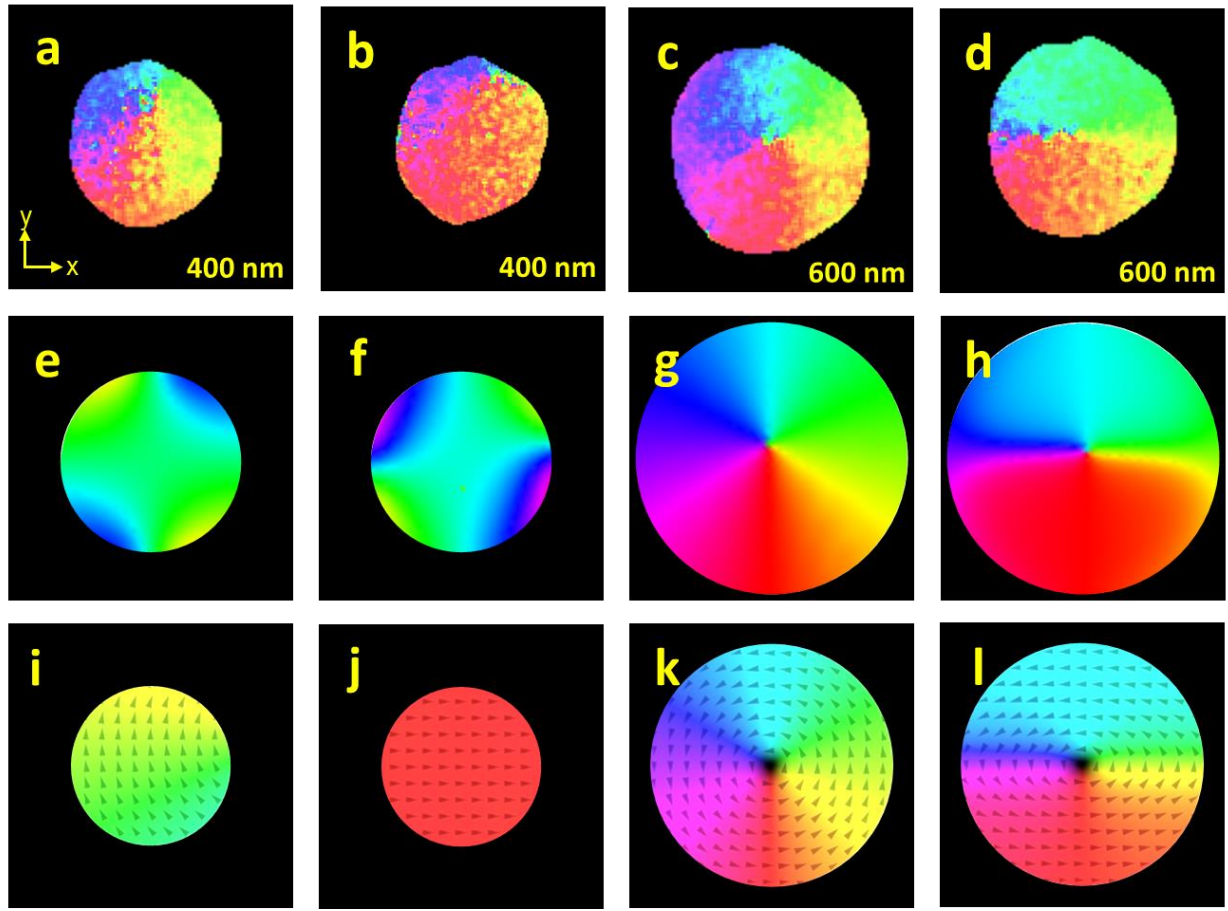

**Figure 3.** The effect of strain on 400 nm and 600 nm diameter Ni disks. Panels (a) and (b) show SEMPA images of the magnetization of a 400 nm Ni disk before and during the application of a 0.8 MV/m electric field to the substrate. The off-center vortex is removed and the magnetization is mostly uniform in this case. Panels (c) and (d) show analogous SEMPA images for a 600 nm Ni disk. In this case, the vortex is compressed by the uniaxial anisotropy induced by the strain into two antiparallel domains. Panels (e) and (f) show the results of the elastodynamic-LLG simulations of the magnetization of a 400 nm Ni disks without (e) and with (f) strain. The strain rotates the magnetization by 90°. Analogous results for the 600 nm disk are presented in (g) and (h). The initial vortex magnetization configuration is compressed into two anti-parallel domains. Panels (i)-(l) show simulations of the same systems as (e)-(h), this time modeled with MuMax3. The color scale used here is the same as that in Figure 2d.

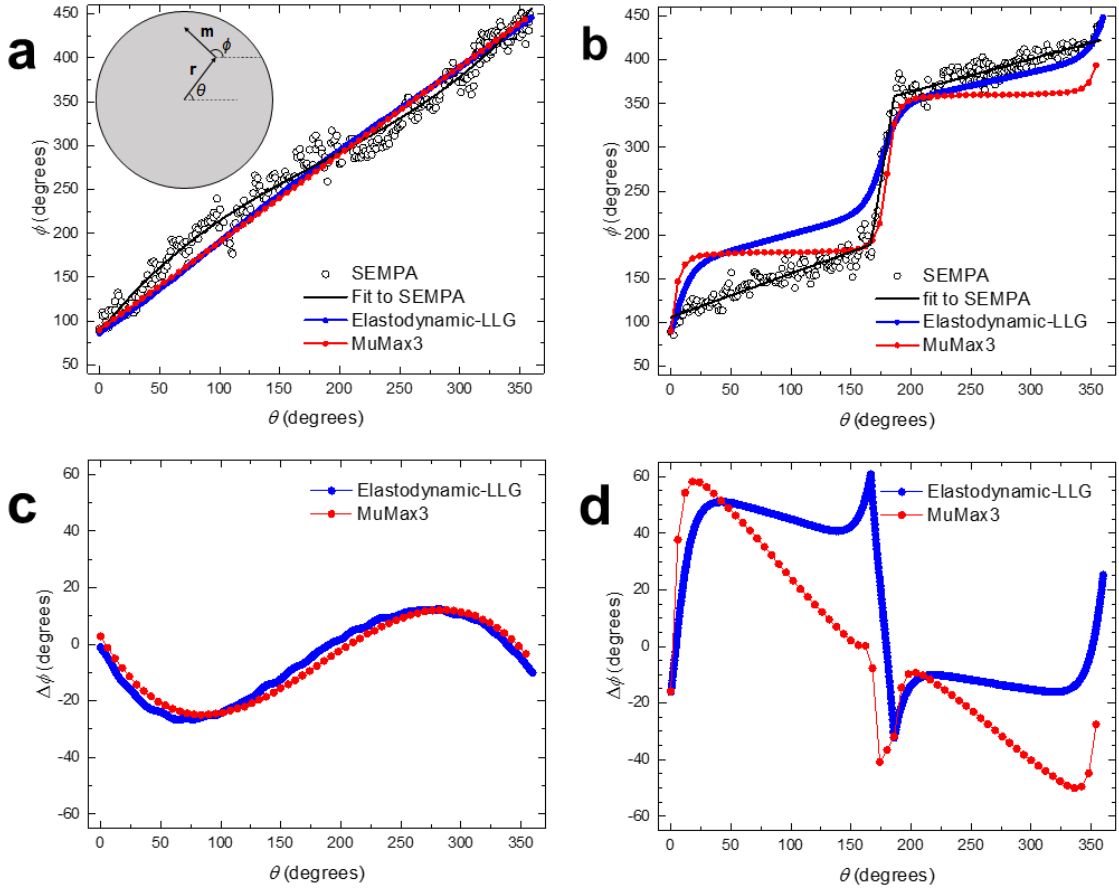

**Figure 4.** Magnetization direction as a function of angular position on the disk for the unstrained (a) and strained (b) 600 nm disks. The inset in (a) defines the angles $\theta$ and $\phi$. The black line in (a) is a 4th order polynomial fit to the SEMPA data, and the black line in (b) is a piecewise linear fit to the SEMPA data. Panels (c) and (d) show the differences between the two simulations and the fits to the SEMPA data for the unstrained and strained cases, respectively. In (d), most of the error for the elastodynamic-LLG simulation appears to be due to the offset between $\theta = 0°$ and $\theta = 180°$, which, as noted in the text, is probably due to the irregular edge of the disk. The SEMPA measurements have an uncertainty of ±6° associated with the counting statistics of the detectors. The simulations agree very well in (c) because in this case, there is no strain, and the coupled elastodynamic-LLG simulation reduces to the same type of micromagnetic simulation as MuMax3.